\documentclass[useAMS,usenatbib,onecolumn,usegraphicx]{mn2e}
\usepackage{color}
\newcommand{\beq}{\begin{equation}}
\newcommand{\eeq}{\end{equation}}
\newcommand{\beqar}{\begin{eqnarray}}
\newcommand{\eeqar}{\end{eqnarray}}
\newcommand{\beqars}{\begin{eqnarray*}}
\newcommand{\eeqars}{\end{eqnarray*}}
\newcommand{\bc}{\begin{center}}
\newcommand{\ec}{\end{center}}
\newcommand{\ben}{\begin{enumerate}}
\newcommand{\een}{\end{enumerate}}
\newcommand{\bit}{\begin{itemize}}
\newcommand{\eit}{\end{itemize}}

\newcommand{\I}{\mbox{\rm i}}

\def \eps{\epsilon}

\renewcommand{\(}{\left(}
\renewcommand{\)}{\right)}

\title[Torsional Oscillations of Slowly Rotating Relativistic Stars]
{Torsional Oscillations of Slowly Rotating Relativistic Stars}
\author[M.~Vavoulidis, A.~Stavridis, K.~D.~Kokkotas and H.~Beyer
] {M.~Vavoulidis$^{1,2}$\thanks{E-mail: miltos@astro.auth.gr},
A.~Stavridis$^{1}$
,
K.D.~Kokkotas$^{1}$
and
H.~Beyer$^{3,4}$
\\
  $^1$ Department of Physics, Aristotle University of Thessaloniki,
  Thessaloniki 54124, Greece \\
  $^2$ Theoretical Astrophysics, University of T\"ubingen, Auf der
  Morgenstelle 10, 72076, T\"ubingen, Germany \\
  $^3$ Louisiana State University (LSU), CCT, 330 Johnston Hall, Baton Rouge, LA 70803, USA \\
  $^4$ Max Planck Institute for Gravitational Physics (AEI),
Am M\"{u}hlenberg 1, 14476 Golm, Germany
}

\begin{document}

\maketitle

\label{firstpage}

\begin{abstract}
We study the effects of rotation on the torsional modes of
oscillating relativistic stars with a solid crust. Earlier works in Newtonian theory
provided estimates of the rotational corrections 
for the torsional modes  and suggested that
they should become CFS unstable, even for quite low rotation rates.
In this work, we study the effect of rotation in the context of general relativity
using elasticity theory and in the slow-rotation approximation.
We find that the Newtonian picture does not change considerably. 
The inclusion of relativistic effects leads only 
to quantitative corrections. 
The degeneracy of modes for different values of $m$
is removed, and modes with $\ell=m$ are 
shifted towards zero frequencies and become secularly unstable
at stellar rotational frequencies $\sim$ 20-30 Hz.
\end{abstract}


\section{Introduction}

Neutron stars are objects of extremely rich internal structure.
Although their interior structure is still very uncertain, it seems
that observations and theoretical studies of neutron stars are quite
in agreement concerning the structure of their exterior parts. More
specifically, there is agreement that neutron stars, 1-2 minutes after
their formation, are cold enough to solidify their exteriors and
form a crystal crust thanks to Coulomb forces between the various
atomic nuclei.
The crust is covered by a very thin fluid ocean, while the interior
is formed by a super-fluid mantle (up to 5km in size). The composition of the core
is highly uncertain. The crystal crust extends from the
neutron star's atmosphere $1$km down where
densities reach nuclear densities around $1.3-2.4\times 10^{14}$
gr/cm$^3$. The Coulomb forces of the crystal ions forming the crust
can be described via the shear modulus $\mu$ which is inversely
proportional to the 4th power of the ion spacing.  Since the 
restoring force is the Coulomb force, the periods of the torsional modes
will strongly depend on the shear modulus, and its values
will characterize the spectrum. Up to now, there are only two detailed
calculations of the  crust equation of state  \citep{NV1973,
DH2001}. 
Both were based on an approximate formulation \citep{Strohmayer1991b}
leading to a typical value for the shear modulus of 
$\mu= 1
\times 10^{30} \mbox{ erg cm}^{-3} \rho_{14}$ \citep{Schumaker1983}.

In the Newtonian limit and in the absence of strong magnetic fields,
\citet{Hansen1980} found that the period of the fundamental
torsional modes $_\ell P_0$ depends mainly on the radius of the
star $R$, the speed of the shear waves $v_s$ and the angular index
$\ell$ via the following relation
\begin{equation}
_\ell \sigma_0 \approx [\ell(\ell+1)]^{1/2} v_s/R 
\label{eq:sigma}
\end{equation}
where  $v_s=(\mu/\rho)^{1/2}$, and $\mu$, $\rho$ are the
shear modulus and the density, respectively.
Torsional modes, which are axial-type oscillations, are
thought to be more easily excited during a fracturing of the crust
since they only involve oscillations of the velocity.  The velocity
field of torsional oscillations is actually divergence-free 
without a
radial component. Torsional modes ({\em t}-modes) are labeled as
$_\ell t_n$, where $\ell$ is the angular index, while the index $n$
corresponds to the number of radial nodes in the eigenfunctions of the
overtones for a specific $\ell$.

The shear and torsional modes are well studied in Newtonian
theory,  see e.g.
\citet{Hansen1980,McDermott1988,Carroll1986,Strohmayer1991a}, while
there are only a few studies in general relativity
\citep{Schumaker1983, Leins1994,Messios2001,SKS2006,SA2007,SKSV2006}.
Relativistic effects have been found to increase significantly the
fundamental $\ell=2$ torsional mode period by roughly 30\% for a
typical $M=1.4M_\odot$, $R=10{\rm km}$ model.

The aim of this work is the study of the effect of rotation on torsional
modes. Up to now this is studied only in Newtonian theory
in a single study by \citet{Strohmayer1991a}, see
also \citet{LS1996}. That study suggested that the
frequency of a torsional mode in a rotating star is given by
\begin{equation}
\sigma=\sigma_0 +\frac{m\Omega}{\ell(\ell+1)}
\end{equation}
where
$\sigma_0$ is the fundamental frequency of the torsional mode of 
a non-rotating star and $\Omega$ is its rotational frequency.
Since the typical frequency of the torsional modes of
non-rotating stars $\sigma_0$ is of the order of 25-40 Hz, it is obvious that
they will be subject to the so called Chandrasekhar-Friedman-Schutz
(CFS) instability, even for small rotational frequencies
\citep{YL2001}. This is a quite interesting result
for rotating neutron stars since the instability (together with that of 
the r-modes) might lead to further fracturing and/or melting of the
crust.

Recently, there is increased interest in the study of torsional modes
because of the belief that Soft Gamma Repeaters (SGRs) could be
magnetars experiencing star-quakes that are connected (through the
intense magnetic field) to gamma ray flare activity. Magnetar
star-quakes may be driven by the evolving intense magnetic field
which accumulates stress and eventually leads to crust fracturing.
There are three SGR events detected up to now. 
The first event was
detected already in 1979 from the source SGR 0526-66
\citep{Mazets1979,Barat1983}, the second in 1998 from SGR 1900+14
\citep{Hurley1999}, while the third and most energetic one was
observed in December 2004 from the source SGR 1806-20
\citep{Terasawa2005, Palmer2005}. Analysis of the tail oscillations,
for a full discussion see \citet{SKS2006,Watts2006}, revealed the presence of
oscillations from a few tenths of Hz up to about 2 kHz. In an
attempt to fit the observed frequencies to the torsional modes of
various EoS for the core and the crust, \citet{SKS2006} suggested
that neutron star models should have stiff equations of state and
a mass between 1.6 - 2 $M_\odot$. Still these results depend critically
on the interpretation of the order of each mode. For this, 
see the discussion in 
\citep{SA2007}.

In the present work, we derive the perturbation equations for
torsional oscillations of rotating stars in general relativity
following the approach by \citet{Kojima1992,SK2005} in the 
Cowling approximation which neglects perturbations of 
the space-time. The article is
organized as follows. In the next section, we describe the
general-relativistic equations that have been used to describe the
background stellar configuration, and we derive the perturbation
equations for torsional oscillations of rotating stars in the
Cowling approximation. In the third section, we describe the
numerical techniques used to calculate the modes
together with a toy
problem. The last is meant to explain the numerical results.
The article closes
with a summary and discussion. 

\section{Perturbation equations for torsional oscillations}

We consider a slowly rotating relativistic star described by the metric
\begin{equation}
ds^2 = -e^{2\nu} dt^2 + e^{2\lambda} dr^2 + r^2 ( d\theta^2 +
\sin^2\theta d\phi^2 ) - 2 \omega r^2 \sin^2\theta dt d\phi
\label{metric}
\end{equation}
where $\nu$, $\lambda$ and $\omega$ (the frequency of the local
frame) are functions of the radial coordinate $r$. Up to first 
order in $\Omega$, the background
4-velocity of the star is given by
\begin{equation}
u^{\mu} = [e^{-\nu},0,0,\Omega e^{-\nu}] \label{Bvelocity}
\end{equation}
where $\Omega$ is the angular velocity of the star. 
The background stellar models are solutions of the TOV 
equations and 
an equation describing the dragging of inertial frames
\begin{equation}
\varpi'' - \left( \nu' + \lambda' - {4 \over r} \right) \varpi' - 16
\pi e^{2\lambda} \left( p + \eps \right) \varpi = 0 \label{varpi}
\end{equation}
where $\varpi := \Omega - \omega$. We assume that the star consists
of a perfect fluid described by the energy-momentum tensor
\begin{equation}
T_{\mu\nu} = (p+\eps) u_{\mu} u_{\nu} + p g_{\mu\nu} \, \, . \label{Tmn}
\end{equation}
We also assume that the star is isotropic,
therefore, the background shear tensor vanishes.

Due to the spherical symmetry of the
background, the perturbations of the background configuration can be decomposed
into spherical harmonics. This leads to a large system of partial differential
equations \citet{Kojima1992,SK2005}. Here the space-time
perturbations are omitted (Cowling approximation) and also the
coupling between spheroidal (polar) and toroidal (axial)
perturbations since they only marginally affect the eigenfrequencies
of the torsional modes.

Under these approximations, the radial component of the perturbed
velocity field and the variations of pressure and density will remain
unaffected since they are polar perturbations.
The perturbation
of the 4-velocity $\delta u^\alpha$ is related to the displacement vector
$\zeta^a$ through the relation $\delta u^\alpha = {\cal L}_u \zeta^\alpha$. For an observer
co-rotating with the star, this is translated to:
\begin{eqnarray}
\delta u^{\theta} &=& 
-e^{-\nu} \dot{Z}
{1 \over \sin\theta} {\partial Y_{\ell m} \over \partial \phi}
\, \, ,
\label{du_theta}
\\
\delta u^{\phi}   &=& 
e^{-\nu} \dot{Z}
{1 \over \sin\theta} {\partial Y_{\ell m} \over \partial \theta}
\, \, , 
\label{du_phi}
\\
\delta u^t &=&
(\Omega - \omega) r^2
e^{-3\nu} \dot{Z} \sin\theta {\partial Y_{\ell m} \over \partial \theta}
\, \, .
\label{du_t}
\end{eqnarray}
In other words, the velocity perturbations are described by  
the time derivative a function $Z=Z(t,r)$ related to the displacement vector $\zeta^a$, here the dot stands for the temporal derivative. The
perturbed energy-momentum tensor, which includes 
the contribution from shear, is given by 
\begin{equation}
\delta T_{\mu\nu} = ( p + \eps ) (
\delta u_{\mu} u_{\nu} + u_{\mu} \delta u_{\nu} ) - 2 \mu \delta S_{\mu\nu}
\label{dTmn}
\end{equation}
where $S_{\mu\nu}$ is the shear tensor, $S_{\mu\nu}$ is
defined by ${ \sigma_{\mu \nu}} = {\cal L}_{u} { S_{\mu \nu}} $. Here
$\sigma_{\mu\nu}$ is the rate of shear given by the Lie derivative
of the shear along the world lines \citep{CQ1972}
\begin{equation}
\sigma_{\mu\nu} = {1 \over 2} \( u_{\mu \, ; \alpha}
P^{\alpha}_{\;\; \nu} +  u_{\nu \, ; \alpha} P^{\alpha}_{\;\; \mu}
\) - { 1 \over 3} P_{\mu\nu} u^{\beta}_{\;\; ;\beta} \label{sigma} \, \, ,
\end{equation}
and $P_{\mu\nu}$ is the projection tensor
\begin{equation}
 P_{\mu\nu} = g_{\mu\nu} + u_{\mu} u_{\nu} \, \, . \label{Pab}
\end{equation}
The speed of shear waves on the crust depends on the shear modulus
$\mu$, the density $\eps$ and the 
pressure $p$ of the star according to $v_s^2 =
\mu/(p+\eps)$. A typical value of the speed of shear waves is $v_s
\approx 10^8 {\rm cm/s}$.  The perturbation equation for the 
energy-momentum tensor $\delta
(T_{\mu\nu}^{\;\;\;\; ; \mu})  =  0$ provides a single equation
for the axial perturbations

\begin{eqnarray}
 \label{eq:t_master}
\ddot{Z} &=& v_s^2 e^{2\nu-2\lambda} \left[Z'' +
\left( {4 \over r} + \nu' -\lambda'
+ \frac{\mu'}{\mu} \right) Z'-
e^{2\lambda} {\Lambda - 2 \over r^2} Z \right]
+ 2 \I m \varpi \left[{1 \over \Lambda} +
v_s^2 \left( 1 - {2 \over \Lambda} \right)
\right] \dot{Z}
\end{eqnarray}
where $\Lambda=\ell(\ell+1)$.
In the absence of shear $\left(\mu=0, v_s^2=0\right)$,
an inertial observer obtains
\begin{equation}
\sigma=- m\left(\Omega-\frac{2\varpi}{\Lambda}\right) \label{eqU2}
\, \, .
\end{equation}
This gives the well known relation for the r-mode frequency in the 
Newtonian limit ($\Omega=\varpi $),
while in the relativistic case it leads to a continuous
spectrum \citep{Kojima1998,BK1999}. In the last case couplings
between higher $l$ and polar perturbations need to be included 
for a proper study of the spectrum.

\section{Estimates of the mode frequencies}

In this section, we present an approximate solution of 
the boundary value problem and describe the numerical techniques used in the calculation of the
frequencies of torsional modes of rotating stars.

For the numerical estimation of the frequencies, we use 
two different techniques. 
The first approach assumes a harmonic time-dependence 
of the perturbations. This leads on a boundary value problem.
The second approach is based on a direct time
evolution of equation (\ref{eq:t_master}) followed by a Fourier transform
in time of the obtained values at a fixed radial position.
We will describe only the first approach. The second approach 
has only been used for the verification 
of the results.

The Fourier transform of equation (\ref{eq:t_master}) ,i.e., the assumption
that $\dot{Z} = \I \sigma Z$, leads to the following differential
equation  
\begin{eqnarray}
 \label{eqUe0} Z'' +  \left( {4 \over r} + \nu'
-\lambda' + \frac{\mu'}{\mu} \right) Z'
+ e^{2\lambda-2\nu} \left\{\frac{\sigma_r^2}{v_s^2} - 2 m
\frac{\varpi}{v_s} \frac{\sigma_r}{v_s} \left[
{1 \over \Lambda} + v_s^2 \left(
1 - {2 \over \Lambda} \right) \right]
- e^{2\nu} {\Lambda - 2\over r^2}
\right\} Z = 0 \, \, .
\end{eqnarray}
The boundary conditions in the center (or at the lower end
of the crust) and on the stellar surface are
\begin{equation}
Z \sim r^{\ell-1} \quad \mbox{for} \quad r\rightarrow 0 \quad \mbox {or}
\quad Z'=0 \quad \mbox{for} \quad r\rightarrow R_c \quad \mbox{and} \quad
Z'=0 \quad \mbox{for} \quad r\rightarrow R
\label{eq:bc1}
\end{equation}
where $R_c$ is the distance of the lower end of the crust from the center.

The above system of equations defines an eigenvalue problem for the frequencies 
of the torsional modes $\sigma$. For its solution, we use two approaches. The 
first is an approximate analytic solution, and the second is a numerical solution. 

\subsection{Approximate solution}

Here we provide an approximate analytic solution to the eigenvalue problem  
based on Bessel functions. In order to be able to use Bessel functions and to 
treat the problem semi-analytically, we need to simplify equation (\ref{eqUe0}).
For this we  
make the following simplifying assumptions, $e^\nu=e^\lambda \approx 1$,
$\mu'=\nu'=\lambda' \approx 0$, $\varpi=\Omega$ and $v_s^2 \ll 1$. This  reduces 
equation  (\ref{eqUe0})  to a Bessel equation
\begin{equation}
Z''+  \frac{4}{r} Z' + \left[\Sigma^2 -\frac{(\ell-1)(\ell+2)}{r^2} \right] Z = 0 
\label{eq:toy4}
\end{equation}
%
where
\begin{equation}
\Sigma^2  =
\frac{\sigma_r^2}{v_s^2}
- \frac{2 m}{\Lambda}
\frac{\Omega}{v_s}
\frac{\sigma_r}{v_s}
\, \, .
\label{eq:toy3}
\end{equation}
Together with the boundary conditions (\ref{eq:bc1}), this equation leads 
on a simpler boundary value problem. Actually, for high frequencies, 
the eigenvalues  could easily be estimated using the WKB method. 
On the other hand,  
since the frequency of the fundamental torsional mode is relatively small, 
it is unclear whether the WKB approximation is applicable, in particular 
in the discussion of  CFS instability. The last involves 
the investigation 
of the limit $\Sigma \to 0$.

The general solution of equation (\ref{eq:toy4}) can be given in the
form of Bessel functions
\begin{equation}
Z(r)= c_1 r^{-3/2} J_{\ell+\frac{1}{2}}\left(\Sigma \, r \right)  +
c_2 r^{-3/2} Y_{\ell+\frac{1}{2}}\left( \Sigma \, r \right) \, .
\label{eq:Bessel}
\end{equation}
where $c_1$ and $c_2$ are arbitrary constants.
Since the condition for regularity in the center demands that $Z\sim
r^{l-1}$, the contribution of the Bessel functions $Y_n(r)$ is
excluded since it is divergent for $r\to 0$. The second boundary
condition demands that $Z'(R)=0$ which leads to the following
transcendental equation
\begin{equation}
(\ell-1)J_{\ell+\frac{1}{2}}
 \left(k \right)- k \, J_{\ell+\frac{3}{2}}\left(k \right)=0,
\end{equation}
where $k=\Sigma \cdot R$. The roots of this equation can be found
by elementary numerical methods. This leads to the following values for $k$  
\begin{eqnarray}
\ell=2 &:& {_2k_n}  =2.501, \, 7.136, \, 10.515, \, 13.772, \, \dots \\
\ell=3 &:& {_3k_n}  =3.865, \, 8.445, \, 11.882, \, 15.175, \, \dots \\
\ell=4 &:& {_4k_n}  =5.095, \, 9.713, \, 13.211, \, 16.544, \, \dots \\
\cdots \, \,  . \nonumber
\end{eqnarray}
The
eigenfrequencies can be obtained from the roots of the equation
\begin{equation}
{_\ell\sigma_n}^2 - \frac{2 m \Omega}{\Lambda}  {_\ell\sigma_n}
= \left(\frac{{_\ell k_n}}{R} v_s\right)^2 \, \, \label{eq:toy3a}
\end{equation}
by setting  ${_\ell\sigma_n}^{(0)}={_\ell k_n v_s}/R $ 
(the frequency in the 
absence of rotation).
In this way, we get an approximate form of the torsional mode frequency  $\sigma_r$  
in the rotating frame
\begin{equation}
{_\ell\sigma_n}={_\ell\sigma_n}^{(0)}
\sqrt{1+\left[\frac{1}{{_\ell\sigma_n}^{(0)}}
\frac{m \Omega}{\ell\left(\ell+1\right)}
\right]^2}
+ \frac{m \Omega}{\ell \left(\ell+1\right)}
\,
\label{eq:toy3c}
\end{equation}
which for an inertial observer ($\sigma=\sigma_r - m \Omega$)
has the form
\begin{equation}
{_\ell\sigma_n}={_\ell\sigma_n}^{(0)}
\sqrt{1+\left[\frac{1}{{_\ell\sigma_n}^{(0)}}
\frac{m \Omega}{\ell\left(\ell+1\right)}
\right]^2}
- \frac{m \Omega \left(\ell^2+\ell-1\right)}{\ell \left(\ell+1\right)}
\, \, .
\label{eq:toy3d}
\end{equation}
Both relations agree with the Newtonian results of \citet{Strohmayer1991a}, 
equation (\ref{eq:toy3c}) and of \citet{YL2001} equation (\ref{eq:toy3d}).
Note that $_\ell k_0 \approx \sqrt{\ell(\ell+1)}$ leads to a well known form of the
frequency of the fundamental torsional mode for non-rotating Newtonian stars, see
equation (\ref{eq:sigma}).

\begin{figure}
\includegraphics[width=100mm]{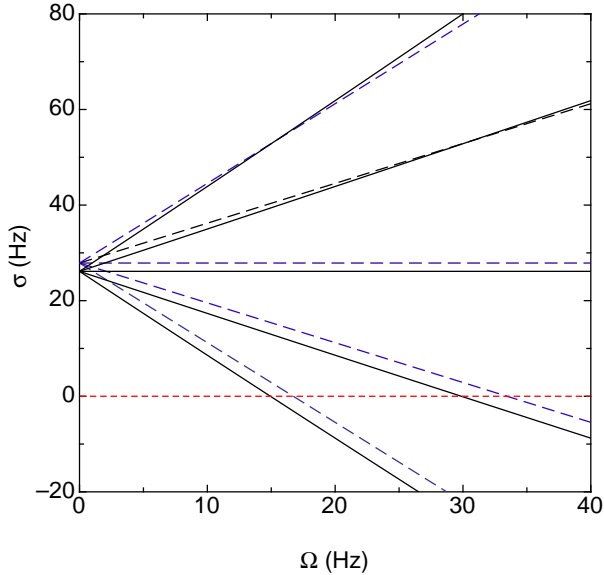}
 \caption{The frequency of the torsional mode $_2t_0$
 as function of the rotation (both in Hz).
 The dashed line corresponds to the frequency of the toy model
 and the continuous line to the frequency of the relativistic model achieved
 by a direct numerical solution of the eigenvalue problem.
The data are for a stellar model with a realistic EoS A \citep{EOS_A} for the fluid core
and an EoS for the crust given by \citet{DH2001} (model $A+DH_{14}$). 
This stellar model has radius $R=9.49$ km $M=1.4 M_\odot$ and crust thickness $\Delta r/R \approx 5.15$\%.
    }
  \label{Fig:1}
\end{figure}

\subsection{Numerical solution}

The numerical solution of the eigenvalue problem defined by equations (\ref{eqUe0}) - (\ref{eq:bc1}) is solved by a shooting method. Its results have been tested by direct 
numerical evolution of the time dependent equation (\ref{eq:t_master}).
The numerical results verify the suggestion by \citet{YL2001} 
that the torsional modes \emph{are CFS unstable}. 

In Figure 1, we plot the frequency of the fundamental torsional mode
($_2t_0$  for $m=-2...2$) as a function of the stellar rotation frequency 
$\Omega$. We also show the results derived by the approximate analytic method 
described earlier.
It is clear that there is a difference in the results of the order of 30\% . 
Mainly, this is due to the fact that in one case we used the 
approximate Newtonian form of the equation,  while in the other
case we used  
its exact relativistic form.

It is obvious that the torsional modes become secularly (CFS) unstable
 even for very slowly rotating relativistic stars. 
In a Newtonian study
\citep{Strohmayer1991a} torsional mode frequencies, measured in a co-rotating reference frame, can be described approximately by
the relation (\ref{eq:sigma}), where $\sigma=\sigma_r$. 
%
Moreover, an inertial observer will measure
\begin{equation}
\sigma_i = \sigma_r - m \Omega \approx \sigma_0 - m \Omega + \frac{m \Omega}{\ell \left(\ell+1\right)} \, \, .
\end{equation}
The CFS instability sets in when the frequency of the 
torsional mode  for  the inertial observer $\sigma_i=0$ or, equivalently, when the phase velocity of the mode is equal to the rotational frequency  $\sigma_r/m = \Omega$, i.e. , when the critical rotation frequency of the star is
given by
\begin{equation}
\Omega_{\mbox{inst}} = \frac{\ell\left(\ell+1\right)}{m \left(\ell^2+\ell-1\right)} \sigma_0 \, .
\end{equation}
The approximate results  derived earlier (\ref{eq:toy3c})-(\ref{eq:toy3d}) verify the Newtonian results. Also the numerical 
results agree extremely well on the effect of rotation on 
the frequencies of the torsional modes, i.e., 
\begin{equation}
\left(\frac{\partial \sigma_i}{\partial \Omega}\right)_{Numerical}\approx
\left(\frac{\partial \sigma_i}{\partial \Omega}\right)_{Approximate} \, 
\end{equation} 
with a typical error of the order of 2-5\%, depending on the compactness of the star.

The perturbation equations have been solved for a number of different equations of state for the fluid core and the crust which 
are listed in Table 1 of \citet{SKS2006}. The overall picture is the same,
i.e., shows a 20-30\% 
difference of the frequencies of the relativistic and the 
Newtonian equations, 
and a 2-5\% difference in
the rate of the frequency as function of rotation 
($\frac{\partial \sigma_i}{\partial \Omega}$). The frequencies plotted in Figure 1 are for a 
stellar model with a realistic EoS A \citep{EOS_A} for the fluid core
together with an EoS for the
crust given by \citet{DH2001} (model $A+DH_{14}$). 
This stellar model has radius $R=9.49$ km $M=1.4 M_\odot$ and crust thickness 
$\Delta r/R \approx 5.15$\%.

\section{Results and Discussion}

In this article, we showed by means of numerical and semi-analytic
methods that the torsional modes of rotating relativistic stars are subject 
to the CFS instability as suggested by \citet{YL2001}
for Newtonian stars.  This instability might be only of academic interest since 
viscosity works against it, and therefore it will probably never prevail. 

We would like to emphasize that the CFS instability does not operate in
the up to now observed SGRs because they are very slowly rotating stars with 
periods of the order of seconds.
Still it is possible that the torsional modes of the newly born neutron stars (soon after 
they form a crust) will be CFS unstable because rotation periods of the order of 
tenths or hundreds of Hz are expected.  The instability of torsional modes in the 
crust and the unstable r-modes in the fluid core might work together in the direction of 
breaking or melting the crust. If strong magnetic fields are present,  then the 
accumulated stress might enhance the above scenario suggesting that the young 
neutron stars with strong magnetic fields will probably have a more frequent flare 
activity. CFS type rotational instabilities of magnetic field modes will also be present in 
rotating neutron stars, and their effect in the stellar flare activity is part of an extension 
of this work.

Finally, rotation (as well as the presence of strong magnetic fields) will produce frequency 
shifts towards both lower and higher frequencies which poses extra difficulties for
the identification of the various observed frequencies from SGRs.

\section*{Acknowledgments}
We are grateful to H.~Sotani, N.~Stergioulas, S.~Yoshida and
J.L.~Friedman for helpful discussions. This work is supported by the
Greek GSRT Programs Heracleitus and Pythagoras II and by the German
Science Foundation (DFG), via a SFB/TR7.

\end{document}